\documentclass[12pt]{article}
\DeclareMathAlphabet{\scr}{U}{rsfs}{m}{n}
\usepackage{latexsym}
\usepackage[mathscr]{eucal}
\usepackage{amsfonts}
\usepackage{amscd}
\usepackage{cite}
\usepackage{array}   
\usepackage{amssymb}
\usepackage{colordvi}
\usepackage[centertags]{amsmath}
\usepackage{enumerate}
\usepackage{graphicx}
\usepackage{booktabs}
\usepackage{theorem}
\usepackage[footnotesize]{caption}
\usepackage{soul}

\newcommand{\cleqn}{\setcounter{equation}{0}}
\newcommand{\newc}{\newcommand}
\newc{\be}{\begin{equation}}
\newc{\ee}{\end{equation}}
\newc{\bea}{\begin{eqnarray}}
\newc{\eea}{\end{eqnarray}}
\newc{\ol}{\overline}
\newc{\wt}{\widetilde}
\newc{\bs}{\boldsymbol}
\newc{\m}{\mathcal}
\newc{\la}{\langle}
\newc{\ra}{\rangle}

\begin{document}

\title{\hfill ~\\[-30mm]
\hfill\mbox{\small IPPP-11-83}\\[-3.5mm]
\hfill\mbox{\small DCPT-11-166}\\[13mm]
       \textbf{$\bs{A_4}$ models of tri-bimaximal-reactor mixing \\[4mm]}}

\date{}

\date{}
\author{
Stephen~F.~King$^{1\,}$\footnote{E-mail: \texttt{king@soton.ac.uk}}~~and~
Christoph~Luhn$^{1,2\,}$\footnote{E-mail: \texttt{christoph.luhn@durham.ac.uk}}\\[9mm]
{\small\it
$^1$School of Physics and Astronomy,
University of Southampton,}\\
{\small\it Southampton, SO17 1BJ, U.K.}\\[3mm]
{\small\it
$^2$Institute for Particle Physics Phenomenology,
University of Durham,}\\
{\small\it Durham, DH1 3LE, U.K.}
}

\maketitle

\begin{abstract}
\noindent  
Recent results from T2K, MINOS and Double CHOOZ all indicate a sizeable reactor angle $\theta_{13}$
which would rule out {\em conventional} tri-bimaximal lepton mixing.
However, it is possible to maintain the tri-bimaximal solar and atmospheric mixing angle predictions, 
 $\theta_{12}\approx 35^\circ$, $\theta_{23}\approx 45^\circ$ even for a quite sizeable 
reactor angle $\theta_{13}\approx ~8^\circ$, using an ansatz
called tri-bimaximal-reactor (TBR) mixing proposed by one of us some time ago. 
We propose an explicit $A_4$ model of leptons based on the type I seesaw mechanism at both the effective and the 
renormalisable level which, together with vacuum alignment, 
leads to surprisingly accurate TBR neutrino mixing, with the second order
corrections to mixing angles having small coefficients. 
 \end{abstract}
\thispagestyle{empty}
\vfill
\newpage
\setcounter{page}{1}




\section{Introduction}
\cleqn

Recently T2K have published evidence for a 
large non-zero reactor angle \cite{Abe:2011sj} which, when combined with data from MINOS and other
experiments in a global fit yields \cite{Fogli:2011qn},
\be
\theta_{13}=8^\circ\pm 1.5^\circ,
\ee
where the reactor angle is defined in the usual PDG convention \cite{Nakamura:2010zzi} and the
errors indicate the one $\sigma$ range, although the statistical significance of a non-zero
reactor angle is about~3$\sigma$. A non-zero reactor angle is also consistent with the first results from
Double CHOOZ~\cite{DCHOOZ}.

%
If confirmed, these recent observations would rule out {\em conventional}
tri-bimaximal (TB) mixing which predicts a zero reactor
angle~\cite{Harrison:2002er}. Many alternative proposals~\cite{flurry} have
been put forward since the first T2K results, all aiming to accommodate a non-zero
$\theta_{13}$. Typically, such models also yield (sizable) deviations to the other
predictions of TB mixing, namely tri-maximal solar mixing $\theta_{12}=
35.26^\circ$ and maximal atmospheric mixing $\theta_{23}= 45^\circ$.
However, since the latter remain in good agreement with current global fits
\cite{Fogli:2011qn}, it would be nice if, somehow, these good
predictions of TB mixing could be maintained while at the same time allowing a
non-zero reactor angle.
In fact the idea of maintaining the TB predictions for the solar and atmospheric angles,
while switching on the reactor angle, was suggested 
some time ago (before T2K results) by one of us and was
referred to as tri-bimaximal-reactor (TBR) mixing \cite{King:2009qt}.

The TBR ansatz \cite{King:2009qt} postulates
a free reactor angle $\theta_{13}$ but with fixed $s_{12}=1/\sqrt{3}$ and $s_{23}=1/\sqrt{2}$ 
corresponding to tri-maximal solar mixing 
$\theta_{12}= 35.26^\circ$ and maximal atmospheric mixing $\theta_{23}= 45^\circ$.
In the PDG convention for the PMNS mixing matrix \cite{Nakamura:2010zzi} the
TBR ansatz for the mixing matrix is then \cite{King:2009qt},
\begin{eqnarray}
U_{\mathrm{TBR}} =
\left( \begin{array}{ccc}
\sqrt{\frac{2}{3}} c_{13} & \frac{1}{\sqrt{3}} c_{13}& s_{13}e^{-i\delta} \\ 
-\frac{1}{\sqrt{6}}(1+ \sqrt{2}s_{13}e^{i\delta})
& \phantom{-} \frac{1}{\sqrt{3}}(1- \frac{1}{\sqrt{2}}s_{13}e^{i\delta})
& \frac{1}{\sqrt{2}} c_{13}\\
\phantom{-}\frac{1}{\sqrt{6}}(1- \sqrt{2}s_{13}e^{i\delta})
& -\frac{1}{\sqrt{3}}(1+ \frac{1}{\sqrt{2}}s_{13}e^{i\delta})
 &  \frac{1}{\sqrt{2}}c_{13}
\end{array}
\right)P,
\label{TBR}
\end{eqnarray}
where
$P={\rm diag}(1,e^{i\frac{\alpha_{21}}{2} }, e^{i\frac{\alpha_{31}}{2} })$
contains the usual Majorana phases.
Note that TBR mixing reduces to TB mixing in the limit that $\theta_{13} \rightarrow 0$.
However in general TBR mixing involves an arbitrary reactor angle
$\theta_{13}$ which could in principle be large, without causing any deviations from
TB solar and atmospheric mixing $s_{12}=1/\sqrt{3}$ and $s_{23}=1/\sqrt{2}$.

The TBR ansatz in Eq.~\eqref{TBR} is clearly very simple to write down. The obvious question is whether 
there is any model that can give rise to TBR mixing. 
Such a model of TBR mixing should also explain
the smallness of the reactor angle $\theta_{13}$ as
compared to the tri-maximal solar mixing angle
$\theta_{12}= 35.26^\circ$ and the maximal atmospheric mixing angle $\theta_{23}= 45^\circ$.
At the same time that the TBR ansatz was proposed, a mechanism was suggested called 
partially constrained sequential dominance (PCSD) which could lead to TBR mixing \cite{King:2009qt}.
Although no actual model was proposed, it was shown how PCSD could originate due to a distortion
of the vacuum alignment which was previously used to account for TB mixing via
constrained sequential dominance (CSD),
which in turn could originate from $A_4$ family symmetry \cite{King:2009qt}. 

In this paper we propose the first explicit $A_4$ model of leptons based on the type~I seesaw mechanism 
at both the effective and the 
renormalisable level which, together with vacuum alignment, leads to PCSD and hence TBR mixing.
To understand the approach we are following in this paper, it is useful to 
first recall the CSD approach to TB mixing.
The basic ingredients there are sequential dominance (SD)
\cite{King:1998jw,arXiv:1003.5498} and vacuum alignment. The strategy of
combining SD with vacuum alignment is familiar from the CSD 
approach to TB mixing \cite{King:2005bj} where a neutrino mass hierarchy is
assumed and the dominant and subdominant flavons responsible for the
atmospheric and solar neutrino masses are aligned in the directions of the 
second and third columns of the TB mixing matrix, namely 
$ (1,1,-1)^T$ and $(0,1,1)^T$.
The idea of PCSD \cite{King:2009qt} is to simply 
maintain the subdominant flavon alignment,
$ \phi_{\nu_2}  = (1,1,-1)^T$, while considering a small perturbation $\varepsilon$ to 
the dominant flavon alignment, $\phi_{\nu_3} =  (\varepsilon ,1,1)^T$. 
Assuming PCSD, with a strong neutrino mass hierarchy,
$ |m_3| > |m_2| \gg |m_1|\approx 0$, leads to a perturbed neutrino mass matrix $m_{\nu}$ \cite{King:2009qt},
\be
m_{\nu}=  \mbox{$\frac{m_2^0}{3}$} \phi_{\nu_2}{\phi_{\nu_2}}^T + \mbox{$\frac{m_3^0}{2}$} {\phi_{\nu_3}\phi_{\nu_3}}^T,
\label{quadratic}
\ee
where $m_2^0$ and $m_3^0$ are the leading order neutrino mass eigenvalues.
The explicit $A_4$ models we propose involve the accurate flavon alignments 
$ \phi_{\nu_2}  = (1,1,-1)^T$ and $\phi_{\nu_3} =  (\varepsilon ,1,1)^T$ which
appear quadratically and reproduce the neutrino mass matrix as in
Eq.~\eqref{quadratic} quite accurately in the case of the renormalisable model. 

Another important result of our paper concerns the analytic diagonalisation of the neutrino mass matrix
in Eq.~\eqref{quadratic} to second order~$\varepsilon^2$. Previously it has been shown that, 
to leading order in $|m_2|/|m_3|$ and $\varepsilon$, Eq.\eqref{quadratic}
leads to TBR mixing in Eq.~(\ref{TBR}) \cite{King:2009qt,arXiv:1011.6167}. 
Remarkably, we find that the coefficients of the second order corrections to the mixing angles are suppressed, 
making the approximate TBR mixing resulting from  Eq.~\eqref{quadratic} to be 
more accurate than expected. Thus the renormalisable $A_4$ model predicts surprisingly accurate
TBR mixing with $\theta_{12}\approx 35^\circ$, $\theta_{23}\approx 45^\circ$ even for quite sizeable $\theta_{13}$. 
 
We note that PCSD is based on the general approach to model building known as the
{\em indirect} approach \cite{King:2009ap} which involves
the quadratic appearance of the flavons as in Eq.~\eqref{quadratic}. 
Recently an alternative type of model of TBR mixing has been proposed based on  
$S_4$ family symmetry, with a type~I plus type~II seesaw mechanism, although no detailed model was proposed
and the necessary vacuum alignment was not studied \cite{Morisi:2011pm}.  
Moreover, TBR mixing does not follow in a general way from the model in
\cite{Morisi:2011pm}, but only occurs for special choices of parameters.

The layout of the rest of the paper is as follows. In Section \ref{sec-a4model} we 
propose an explicit $A_4$ model of leptons at both the effective and the 
renormalisable level which, together with vacuum alignment, leads to PCSD and hence TBR mixing.
In Section \ref{sec-tbr} we give the results of the analytic diagonalisation of
the neutrino mass matrix in Eq.~\eqref{quadratic} to second
order~$\varepsilon^2$. The straightforward generalisation to the case with a
non-zero lightest neutrino mass $m_1\neq 0$ is presented in~Appendix
\ref{sec-withM1} and might be relevant for some future model. 
Section \ref{sec-concl} concludes the paper. 




\section{\label{sec-a4model}${\bs{A_4}}$ models of tri-bimaximal-reactor mixing}
\cleqn

In this section we present the details of the $A_4$ model, first at the
effective, then at the renormalisable level. As pointed out in the
introduction, our models follow the indirect approach and the aligned flavons appear
quadratically in the light neutrino mass matrix~$m_\nu$ as in Eq.\eqref{quadratic}. 
In order to achieve this via the type I see-saw mechanism, and to ensure a diagonal
charged lepton mass matrix, additional $Z_4$ shaping symmetries are also employed.
In such models, the discrete
family symmetry is pivotal in two aspects: it ($i$) helps generate the
required flavon alignments and ($ii$) combines the three generations into
multiplets of the group so that, together with the corresponding product
rules, one obtains mass matrices which depend on only a few free parameters.
In this paper we choose to work with $A_4$ as it is the smallest non-Abelian
finite group with an irreducible triplet representation. We apply the $A_4$ basis in
which the triplets are explicitly real as given for example
in~\cite{A4-refs}. Denoting a general $A_4$ triplet as
${\bf{c}}=(c_1,c_2,c_3)^T$ and defining $\omega=e^{2\pi i/3}$, the product
rules can be summarised as 
\be
{\bf{c\otimes c'}}  ~=~ \sum_{r=0}^2 (c^{}_1 c'_1 + \omega^{-r} c^{}_2 c'_2 +
\omega^{r} c^{}_3 c'_3 ) \;+\,
\begin{pmatrix} c^{}_2c'_3 \\ c^{}_3c'_1\\c^{}_1c'_2  
\end{pmatrix}
+
\begin{pmatrix} c^{}_3c'_2 \\ c^{}_1c'_3\\c^{}_2c'_1  
\end{pmatrix} \ ,
\ee
 corresponding to two triplets and the sum of the three one-dimensional
 irreducible representations ${\bf{1_r}}$, with ${\bf{1_0=1}}$ being the trivial
 singlet. Furthermore, ${\bf{1_r \otimes 1_{s}= 1_{(r+s) \text{\,mod~3}}}}$.


\subsection{The effective $\bs{A_4}$ model}

The complete list of lepton, Higgs and flavon fields introduced in our model is given in
Table~\ref{tab-A4}. Similar to the $A_4$ model of \cite{Antusch:2011sx} we
have a $U(1)_R$ symmetry as well as several  $Z_4$ shaping symmetries.

\begin{table}
$$
\begin{array}{ccccccccccccccccc}\toprule
&N^c_2&N^c_3&L&e^c&\mu^c & \tau^c
&H_{u}&H_{d}
&\varphi_e&\varphi_\mu&\varphi_\tau
&\varphi_{\nu_2}&\varphi_{\nu_3}&\varphi_a&\varphi_{\tilde a}&\xi\\  
\midrule
A_4 & {\bf 1} & {\bf 1} & {\bf 3} & {\bf 1} & {\bf 1} & {\bf 1} 
& {\bf 1}  & {\bf 1}
&{\bf 3} & {\bf 3} & {\bf 3} 
& {\bf 3} & {\bf 3}& {\bf 3}& {\bf 3}& {\bf 1} 
\\[2mm]
Z_4^{\nu_2} & \eta^3& 1   & 1 & 1&1&\eta
&1&1
&1& 1 &\eta^3 
& \eta&1  &1&1&\eta\\[2mm]
Z_4^{\nu_3} & 1 & \eta^3  & 1 & 1&1&1
&1&1
&1& 1 &1 
&1& \eta  &1&1&1\\[2mm]
Z_4^{e} & 1 &  1 & 1 &\eta^3  &1&\eta^3
&1&1
&\eta &1& \eta
&1& 1  &1&1&1\\[2mm]
Z_4^{\mu} & 1 &  1 & 1 &1&\eta^3  &1
&1&1
&1&\eta &1
&1& 1  &1&1&1\\[2mm]
Z_4^{a} & 1 & 1  & 1 & 1&1&1
&1&1
&1& 1 &1 
&1& 1  &\eta&\eta^2&1\\[2mm]
U(1)_R&1&1&1&1&1&1&0&0&0&0&0&0&0&0&0&0 \\ \bottomrule
\end{array}
$$
\caption{\label{tab-A4}Lepton, Higgs and flavon fields in an indirect $A_4$ model.}
\end{table}

The neutrino part of the effective superpotential reads,
\bea
W_{A_4}^{\nu, \mathrm{eff}} &\sim & \sum_{i=2}^3\left(
LH_u \frac{\varphi_{\nu_i}}{M_{\chi_i}} N_i^c
+ 
N^c_iN^c_i \frac{\varphi_{\nu_i}\varphi_{\nu_i}}{M_{\Upsilon_{i}}} \right)  ,
\label{a4-Ynu0}
\eea
where the effectively allowed mixing term
$N_2^cN_3^c\varphi_{\nu_2}\varphi_{\nu_3}$ must be forbidden by the choice of
appropriate messengers.  Notice that there are only two right-handed neutrinos
living in the singlet representation of $A_4$. Hence, the model will feature
one massless light neutrino.
Inserting the flavon VEVs
\be
\langle \varphi_{\nu_2} \rangle = 
v_{\nu_2} \begin{pmatrix} 
  1\\1\\-1 \end{pmatrix}  , 
\qquad
\langle \varphi_{\nu_3} \rangle =  
v_{\nu_3} \begin{pmatrix} \varepsilon\\1\\1 \end{pmatrix} ,  
\label{a4-align-nu}
\ee
 whose alignment is discussed later, leads to the following Dirac and
 right-handed Majorana neutrino mass matrices
\be
m_D~=~\begin{pmatrix} a_2&a_3\, \varepsilon  \\a_2 & a_3\\-a_2 & a_3 \end{pmatrix} \ ,
\qquad
M_R~=~ \begin{pmatrix} M_2 &0\\0&M_3\end{pmatrix}
\ ,\label{mR}
\ee
where 
$a_i\sim \frac{ v_u v_{\nu_i}}{M_{\chi_i}}$ and $M_i \sim
\frac{v_{\nu_i}^2}{M_{\Upsilon_i}}$. Note that $M_R$ is diagonal by construction.

Using the type~I seesaw formula we can express the light neutrino mass matrix as
\be
m_\nu^{}
= 
m_D^{} M_R^{-1} m_D^T
=  
\frac{a_2^2}{M_2} \begin{pmatrix}  
1 & 1& -1 \\
1&1&-1\\
-1&-1&1
\end{pmatrix}
+
\frac{a_3^2}{M_3} \begin{pmatrix}  
\varepsilon^2 & \varepsilon& \varepsilon \\
\varepsilon&1&1\\
\varepsilon&1&1
\end{pmatrix}  \ .\label{meff}
\ee
With this structure we arrive at TBR mixing in the neutrino sector as will be
shown analytically in Section~\ref{sec-tbr}. 

In order to account for the charged lepton mass hierarchy
we identify the right-handed charged leptons with
$A_4$ singlets, and distinguish them using the $Z_4$ shaping symmetries. 
The resulting effective charged lepton superpotential then takes the form
\be
W_{A_4}^{\ell, \mathrm{eff}} ~\sim ~ \frac{1}{M_{\Omega}} H_d 
\left( L \varphi_\tau  \tau^c + L \varphi_\mu  \mu^c + L \varphi_e  e^c \right)
 \ .\label{a4-yuk-eff}
\ee
Inserting the flavon VEVs
\be
\langle \varphi_\tau \rangle = v_\tau 
\begin{pmatrix} 0\\0\\1 \end{pmatrix} \ , \qquad
\langle \varphi_\mu \rangle =v_\mu 
\begin{pmatrix} 0\\1\\0 \end{pmatrix} \ ,\qquad
\langle \varphi_e \rangle =v_e 
\begin{pmatrix} 1\\0\\0 \end{pmatrix}
 \ ,\label{a4-align-char0}
\ee
whose alignment is discussed later leads to
\be
W_{A_4}^{\ell, \mathrm{eff}} ~\sim ~ \frac{1}{M_{\Omega}} H_d 
\left(v_\tau L_3 \tau^c + v_\mu L_2 \mu^c + v_e L_1 e^c \right)
 \ , \label{hierarchy}
\ee
thus yielding a diagonal charged lepton mass matrix $m_\ell$.
In this model, the hierarchy in the charged leptons remains unaccounted for,
however it is straightforward to implement the Froggatt-Nielsen
mechanism~\cite{CERN-TH-2519}  to cure this. For the purpose of clarity we
will ignore this issue in the following. 


\subsection{The renormalisable $\bs{A_4}$ model}

As emphasised in \cite{Varzielas:2010mp}, any non-renormalisable operator of an
effective superpotential should be understood in terms of a more fundamental
underlying renormalisable theory.  Without such a UV completion of a model,
higher order terms which are allowed by the symmetries may or may not be
present. Thus a purely effective formulation would  leave room for different
physical predictions.
We have constructed a fully renormalisable theory of the lepton sector. The
required messengers  are listed in Table~\ref{tab-A4-3}, while the 
 driving fields which control the alignment of the flavons are presented in Table~\ref{tab-A4-2}.

\begin{table}
$$
\begin{array}{ccccccccccccc}\toprule
&\chi_2^{}&\chi_2^c&\chi_3^{}&\chi_3^c
& \Upsilon_2^{} &  \Upsilon_2^c& \Upsilon_3^{} &  \Upsilon_3^c  
& \Omega& \Omega^c
&\Sigma & \Sigma^c
\\  \midrule
A_4 & {\bf 1} & {\bf 1} & {\bf 1} & {\bf 1} 
& {\bf 1} & {\bf 1} & {\bf 1} & {\bf 1} 
& {\bf 3} & {\bf 3}
& {\bf 3} & {\bf 3}
\\[2mm]
Z_4^{\nu_2} &\eta^3 &\eta& 1 & 1 
&\eta^2 &\eta^2& 1 & 1 
&1&1
& \eta &\eta^3
\\[2mm]
Z_4^{\nu_3} & 1 & 1& \eta^3 & \eta 
& 1 & 1& \eta^2 & \eta^2 
&1&1
& \eta &\eta^3\\[2mm]
Z_4^{e} & 1&1&1&1
& 1&1&1&1
&1&1
&1&1
\\[2mm]
Z_4^{\mu} & 1&1&1&1
& 1&1&1&1
&1&1
&1&1
\\[2mm]
Z_4^{a} & 1&1&1&1
& 1&1&1&1
&1&1
&1&1 
\\[2mm]
U(1)_R&1&1&1&1
&0&2&0&2
&1&1
&0&2 \\ \bottomrule
\end{array}
$$
\caption{\label{tab-A4-3}Messenger fields of the renormalisable $A_4$ model.}
\end{table}

With the particle content and the symmetries specified in Tables~\ref{tab-A4}-\ref{tab-A4-3}, we can replace the effective neutrino superpotential in Eq.~\eqref{a4-Ynu0}
by a renormalisable one which includes the messenger fields,
\bea
W^{\nu}_{A_4} &=&  \sum_{i=2}^3
\left( y_i^{} L \varphi_{\nu_i} \chi_i^{} +  y'_i \chi_i^c N_i^c H_u 
+ 
x_i^{} N_i^cN_i^c\Upsilon_i^{} +  x'_i \Upsilon_i^c \varphi_{\nu_i}^{} \varphi_{\nu_i}^{} 
\right) + \tilde x''_2 \Upsilon_2^c \xi\xi \notag \\
&&+ \, \sum_{i=2}^3 \left( M_{\chi_i} \chi_i^{} \chi_i^c + M_{\Upsilon_i} \Upsilon_i^{} \Upsilon_i^c  \right)
\ . \label{a4-numess} 
\eea
Integrating out the messenger pairs $\chi_i^{},\chi_i^c$ and
$\Upsilon_i^{},\Upsilon_i^c$, the effective
operators of Eq.~\eqref{a4-Ynu0} are uniquely generated. Notice that the
messenger pairs $\Upsilon_i^{}$,$\Upsilon_i^c$ do not lead to the aforementioned
mixing term $N_2^cN_3^c \varphi_{\nu_2}\varphi_{\nu_3}$.

The charged lepton sector is formulated at the renormalisable level
using only one new pair of messengers, $\Omega$ and $\Omega^c$.  
With the particles and symmetries listed in Tables~\ref{tab-A4}~and~\ref{tab-A4-3}
we get the renormalisable superpotential for the charged leptons
\bea
W_{A_4}^{\ell} &\sim& 
LH_d \Omega + \Omega^c \varphi_{\tau} \tau^c 
+  \Omega^c \varphi_{\mu} \mu^c
 + \Omega^c \varphi_e e^c  + M_{\Omega} \Omega \Omega^c 
 \ , \label{s4-yuk}
\eea
where we have suppressed all order one coupling constants. 
Integrating out the messengers, we are led uniquely to $W_{A_4}^{\ell, \mathrm{eff}}$ of Eq.~\eqref{a4-yuk-eff}.


\subsection{\label{sec-s4-vac}Vacuum alignment}

\begin{table}
$$
\begin{array}{cccccccccccccc}\toprule
&A_e^{}&A_\mu^{}&A_\tau^{} & O_{e\mu} & O_{e\tau}  & O_{\mu\tau} 
&A_{\nu_2}
& O_{ea}&   O_{\nu_2 a}
&O_{e \tilde a}&O_{a \tilde a}
&O_{\tilde a \nu_3}
&D
\\  \midrule
A_4 & {\bf 3} & {\bf 3} & {\bf 3} & {\bf 1} & {\bf 1} & {\bf 1} 
& {\bf 3} 
& {\bf 1} & {\bf 1}  
& {\bf 1} & {\bf 1}  
& {\bf 1} 
&{\bf 1}  \\[2mm]
Z_4^{\nu_2} & 1 & 1 & \eta^2 & 1&\eta&\eta
&\eta^2
&1&\eta^3
&1&1
& 1&1\\[2mm]
Z_4^{\nu_3} & 1 & 1 & 1 & 1&1&1
&1
&1&1
&1&1
& \eta^3 &\eta^3\\[2mm]
Z_4^{e} & \eta^2 & 1 & \eta^2 & \eta^3&\eta^2&\eta^3
&1
&\eta^3&1
&\eta^3&1
& 1&\eta^3\\[2mm]
Z_4^{\mu} & 1& \eta^2  & 1 & \eta^3&1&\eta^3
&1
&1&1
&1&1
& 1&1\\[2mm]
Z_4^{a} & 1&1&1&1&1&1
&1
&\eta^3&\eta^3
&\eta^2&\eta
&\eta^2&1\\[2mm]
U(1)_R&2&2&2&2&2&2&2&2&2&2&2&2&2 \\ \bottomrule
\end{array}
$$
\caption{\label{tab-A4-2}Driving fields of the $A_4$ vacuum alignment.}
\end{table}

So far we have only postulated the particular alignments of the neutrino-type
flavons, given in Eq.~\eqref{a4-align-nu}, and the flavons of the charged
lepton sector, Eq.~\eqref{a4-align-char0}. In this subsection we explore the
driving potential and prove that the assumed flavon
alignments can in fact be obtained in a relatively simple and elegant way.

The renormalisable superpotential involving the driving fields
necessary for aligning the neutrino-type flavons is given as
\bea
W_{A_4}^{\mathrm{flavon},\nu}&=& 
A_{\nu_2} (g_1 \varphi_{\nu_2}\varphi_{\nu_2}  
+ g_2 \varphi_{\nu_2} \xi )
\label{a4-flavon-nu}
\\[2mm] &&
+ \,O_{ea} g_3 \varphi_e \varphi_a +  O_{\nu_2 a} g_4 \varphi_{\nu_2} \varphi_a 
+ O_{e\tilde a} g_5 \varphi_e \varphi_{\tilde a} +  O_{a \tilde a} g_6
\varphi_{a} \varphi_{\tilde a} 
\notag \\[2mm] &&
+\,O_{\tilde a \nu_3} g_7 \varphi_{\tilde a} \varphi_{\nu_3}
\notag \\[2mm]&&
+\,D (g_8 \varphi_e \varphi_{\nu_3} + g_9 \varphi_\tau \Sigma) + g'_9\Sigma^c\varphi_{\nu_2}\varphi_{\nu_3}
+  g''_9\Sigma^c \xi \varphi_{\nu_3}  
+M_\Sigma \Sigma \Sigma^c \ .\notag
\eea
The first line of Eq.~\eqref{a4-flavon-nu} produces the vacuum alignment 
$\langle \varphi_{\nu_2} \rangle \propto (1,1,-1)^T$ of Eq.~\eqref{a4-align-nu}
as can be seen from the $F$-term conditions\footnote{We remark that the
  general alignment derived from these $F$-term conditions is $\langle
  \varphi_{\nu_2} \rangle \propto (\pm1 ,\pm 1,\pm1)^T$. One can, however,
  show that all of them are equivalent as the resulting mass matrices $m_\nu$
  will be related by appropriate changes of the unphysical phases
  in the matrix $P'$. Our choice leads to $\delta_{e,\mu,\tau}$ all being zero
  in the limit of vanishing $\varepsilon$, see Eqs.~(\ref{delt-e}-\ref{delt-tau}).}  
\be
2 g_1 \begin{pmatrix}
\langle \varphi_{\nu_2} \rangle_2  \langle \varphi_{\nu_2} \rangle_3 \\
\langle \varphi_{\nu_2} \rangle_3  \langle \varphi_{\nu_2} \rangle_1 \\
\langle \varphi_{\nu_2} \rangle_1  \langle \varphi_{\nu_2} \rangle_2 
\end{pmatrix}
+ g_2 \langle \xi \rangle 
\begin{pmatrix}
\langle \varphi_{\nu_2} \rangle_1\\
\langle \varphi_{\nu_2} \rangle_2\\
\langle \varphi_{\nu_2} \rangle_3
\end{pmatrix} ~=~ 
\begin{pmatrix} 0\\0\\0
\end{pmatrix} . 
\ee
The terms in the second line of Eq.~\eqref{a4-flavon-nu} give rise to
orthogonality conditions which uniquely fix the alignments of the auxiliary
flavon fields $\varphi_a$ and $\varphi_{\tilde a}$,
\bea
\langle \varphi_e \rangle^T \cdot \langle \varphi_a \rangle \,=\, 
\langle \varphi_{\nu_2} \rangle^T \cdot \langle \varphi_a \rangle \,=\, 0 
\quad 
&\rightarrow&
\quad
\langle \varphi_a \rangle \,\propto\, \begin{pmatrix} 0\\1\\1\end{pmatrix} \ , \\
\langle \varphi_e \rangle^T \cdot \langle \varphi_{\tilde a} \rangle \,=\, 
\langle \varphi_{a} \rangle^T \cdot \langle \varphi_{\tilde a} \rangle \,=\, 0 
\quad 
&\rightarrow&
\quad
\langle \varphi_{\tilde a} \rangle \,\propto\, \begin{pmatrix} 0\\1\\-1\end{pmatrix} \ .
\eea
Here we have assumed the flavons $\varphi_e$ and $\varphi_{\nu_2}$ to be already
aligned as in Eqs.~(\ref{a4-align-char0}) and (\ref{a4-align-nu}),
respectively. Finally, the neutrino-type flavon $\varphi_{\nu_3}$ gets aligned
by the remaining terms of Eq.~\eqref{a4-flavon-nu}. A vanishing $F$-term of
the driving field $O_{\tilde a \nu_3}$ requires
\be
\langle \varphi_{\nu_3} \rangle = \begin{pmatrix} n_1 \\ n_2 \\ n_2 \end{pmatrix} ,
\ee
where $n_1$ and $n_2$ are independent parameters. They are further
constrained by the $F$-term condition of the driving field $D$ which -- after
integrating out the messenger pair $\Sigma,\Sigma^c$  and inserting the flavon
VEVs -- reads,
\be
g_8 v_e n_1 - \frac{1}{M_\Sigma} g_9 v_\tau  (g_9' v_{\nu_2} + g_9'' \langle
\xi \rangle ) n_2 ~=~ 0 \ .
\ee
This shows  that $n_1$ is naturally suppressed compared to $n_2$ due to the
messenger mass. With $\varepsilon = \frac{n_1}{n_2}$, we get
\be
\langle \varphi_{\nu_3} \rangle ~=~ v_{\nu_3}\begin{pmatrix} \varepsilon \\1
  \\ 1 \end{pmatrix} ,
\ee
as anticipated in Eq.~\eqref{a4-align-nu}.

Turning to the flavon alignment of the charged lepton sector, the
renormalisable driving superpotential takes the form
\be
W_{A_4}^{\mathrm{flavon},\ell}~\sim~
A_e \varphi_e \varphi_e + A_\mu \varphi_\mu \varphi_\mu 
+ A_\tau \varphi_\tau \varphi_\tau
+O_{e\mu} \varphi_e  \varphi_\mu+ O_{e\tau} \varphi_e  \varphi_\tau
+ O_{\mu\tau} \varphi_\mu  \varphi_\tau \ .  \label{a4-align-charged}
\ee
Here we have suppressed all order one coefficients as they are completely irrelevant.

The triplet driving fields $A_{e,\mu,\tau}$ give rise to flavon alignments
$\langle \varphi_{e,\mu,\tau} \rangle$
with two zero components. The singlet driving fields in turn require
orthogonality among the three flavon VEVs so that we end up with the vacuum
structure of Eq.~\eqref{a4-align-char0}, possibly requiring some redefinition
of the family indices.




\section{\label{sec-tbr}Analytic diagonalisation of the PCSD neutrino mass matrix to second order}
\cleqn

In this section we diagonalise the effective neutrino mass matrix of
Eq.~\eqref{quadratic}, 
\be
m_\nu ~=~ 
\frac{m^0_2}{3} \begin{pmatrix}
1&1&-1 \\
1&1&-1\\
-1 &-1&1
\end{pmatrix}
+
\frac{m^0_3}{2} \begin{pmatrix}
\varepsilon^2&\varepsilon&\varepsilon \\
\varepsilon&1&1\\
\varepsilon &1&1
\end{pmatrix},
\label{m-analytic}
\ee
which arises in the indirect $A_4$ models of Section~\ref{sec-a4model}, see
Eq.~\eqref{meff}.  Numerically the complex parameter 
\be
\varepsilon ~=~ \epsilon \, e^{i \delta^0} \ ,
\ee
$\epsilon,\delta^0 \in \mathbb R$, is assumed to take an absolute value  of
$\epsilon \approx 0.2$. It is therefore reasonable to expand the mixing matrix which
diagonalises $m_\nu$ in powers of $\epsilon$. In the limit of
vanishing~$\epsilon$, the absolute values of the mass parameters   
\be
m^0_i ~=~ |m^0_i| e^{-i \alpha^0_i} \ ,
\ee
correspond to the eigenvalues of $m_\nu$, and the mixing matrix is exactly of
tri-bimaximal form. Switching on $\epsilon$ changes both the masses and the
mixing. In what follows we assume a normal hierarchical neutrino mass
spectrum which allows us to parameterise the masses in Eq.~\eqref{m-analytic} as
\be
|m^0_2|~=~k \,\epsilon \,|m^0_3| \ ,\label{k-param}
\ee
with $k \in \mathbb R$ being an order one coefficient. The neutrino mass
matrix $m_\nu$ of Eq.~\eqref{m-analytic} is now diagonalised by the unitary
mixing matrix $U=P' \,U_{\mathrm{PMNS}}^{}$ such that\footnote{Here we assume
  the convention in  which the effective light   neutrino mass matrix $m_\nu$
  is defined through   the  bilinear $\nu_L \nu_L$   coupling, in contrast to the convention
  $\ol \nu_L \ol \nu_L$ adopted in \cite{King:2009qt}.} 
\be
m_\nu^{\mathrm{diag}} ~=~ U^T m_\nu \,U \ .
\ee
$P'=\mathrm{diag}(e^{i\delta_e}, e^{i\delta_\mu},e^{i\delta_\tau})$ 
is an unphysical phase matrix which is required to bring the PMNS matrix into
PDG form,
\be
U_{\mathrm{PMNS}} ~=~ 
\begin{pmatrix}
1&0&0\\
0&c_{23}&s_{23}\\
0&-s_{23}&c_{23}
\end{pmatrix}
\begin{pmatrix}
c_{13}&0&s_{13}e^{-i\delta}\\
0&1&0\\
-s_{13}e^{i\delta}&0&c_{13}
\end{pmatrix}
\begin{pmatrix}
c_{12}&s_{12}&0\\
-s_{12}&c_{12}&0\\
0&0&1
\end{pmatrix}
 P \ ,
\ee
with $c_{ij}=\cos\theta_{ij}$ and $s_{ij}=\sin\theta_{ij}$. The Majorana
phases are included in the matrix 
$P=\mathrm{diag}(1,e^{i\frac{\alpha_{21}}{2}},e^{i\frac{\alpha_{31}}{2}})$.

We have worked out the mixing matrix $U$ to second order in $\epsilon$ using
the above notation. The results read
\bea
\theta_{12}&=& \arcsin \mbox{$\frac{1}{\sqrt{3}}$} -
\mbox{$\frac{\epsilon^2}{6 \sqrt{2}}$} \ ,  \label{t12}\\[1.5mm]
\theta_{23}&=& \mbox{$\frac{\pi}{4}$}+ \mbox{$\frac{\epsilon^2}{3}$} \,
 k \cos(\alpha^0_2-\alpha^0_3+\delta^0)   \ , \label{t23}\\[1.5mm]
\theta_{13}&=& \mbox{$\frac{\epsilon}{\sqrt{2}}$} +
\mbox{$\frac{\epsilon^2}{3\sqrt{2}}$} \,
k \cos(\alpha^0_2-\alpha^0_3+2\delta^0)   \ , \label{t13}\\[1.5mm]
\delta &=& \delta^0 - \mbox{$\frac{\epsilon}{3}$}\,
 k \sin(\alpha^0_2-\alpha^0_3+2\delta^0)  \ , \label{delt}\\[1.5mm]
\alpha_{21}&=&   \alpha^0_2
\ , \label{al2}\\[1.5mm]
\alpha_{31}&=& \alpha^0_3
\ ,\label{al3}\\[1.5mm]
\delta_e &=&  0 
\ ,\label{delt-e}\\[1.5mm]
\delta_\mu &=&  
\mbox{$\frac{ \epsilon^2}{3}$}\,  k  \sin(\alpha^0_2-\alpha^0_3+\delta^0)
 \ ,\label{delt-mu} \\[1.5mm]
\delta_\tau &=& 
-  \mbox{$\frac{\epsilon^2}{3}$}\, k \sin(\alpha^0_2-\alpha^0_3+\delta^0)\  ,\label{delt-tau}
\eea
leading to a diagonalised mass matrix of the form
\be
m_\nu^{\mathrm{diag}} 
~=~ 
\text{diag} \left(m_1\,,\,m_2\,,\,m_3\right) 
~=~ 
\text{diag} \left(0 \,,\, k \,\epsilon \,,\, 1+
\mbox{$\frac{\epsilon^2}{2}$}\right)  \, |m^0_3|\ .
\ee
Notice that the Dirac CP phase  $\delta$ is only given to first order in
$\epsilon$  as it always appears together with $\sin\theta_{13}$ whose leading
term is already suppressed by one power of $\epsilon$. 
The Majorana phases on the other hand do not receive any corrections of order
$\epsilon$ or $\epsilon^2$. With $m_1=0$  there is only one
physical Majorana phase,
\be
\alpha_{23} ~=~ \alpha_{21} - \alpha_{31} ~=~ \alpha^0_2 - \alpha^0_3\ .
\ee
To first order in 
$\epsilon$ the deviations in the solar and atmospheric mixing vanish, and only
the reactor angle $\theta_{13}$ is switched on. Hence the mass matrix of
Eq.~\eqref{m-analytic} gives rise to TBR mixing at leading order.
In more detail, the second order results show that the solar mixing angle is corrected to
values slightly smaller than the TB value, while the atmospheric angle can
deviate to larger or smaller values, depending on the phases. Furthermore, the
deviation of $\theta_{23}$ is bigger by a factor of about $2\sqrt{2}$ compared to
the deviation of $\theta_{12}$.

\begin{figure}
\centering
\includegraphics[scale=0.71]{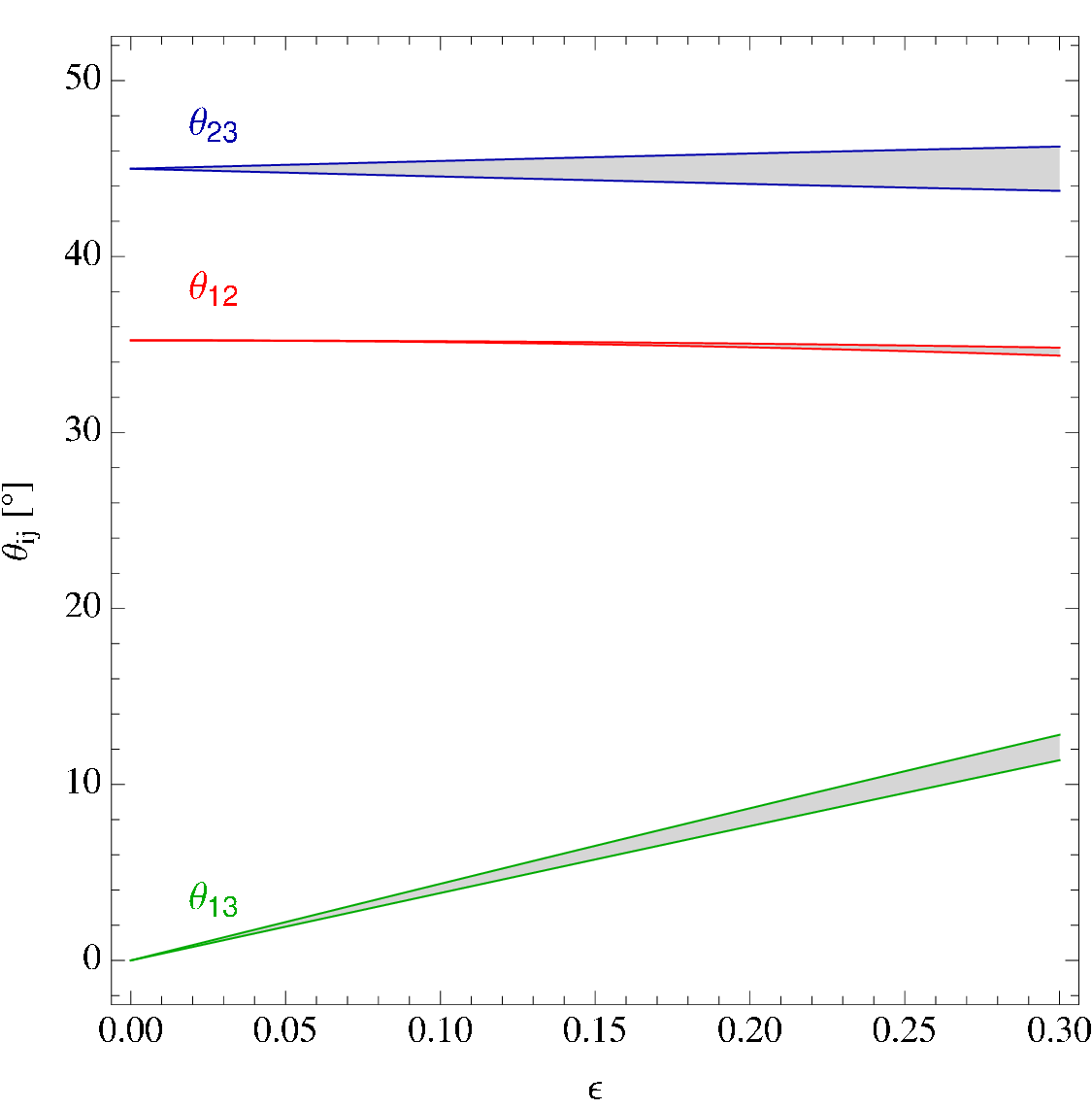}
\caption{\label{antusch-plot}The change of the solar, atmospheric and reactor mixing
  angles with increasing $\epsilon$ in PCSD. This plot, which 
 first appeared without the phase dependence in \cite{arXiv:1003.5498}, 
 was obtained using
  the Mixing Parameter Tools of the REAP Mathematica
  package~\cite{hep-ph/0501272}.
  The numerical results shown are consistent with our analytic second order
  equations as discussed   in the text.}
\end{figure}

Our analytic expressions confirm the numerical results for PCSD obtained with
the Mixing Parameter Tools of the REAP Mathematica
package~\cite{hep-ph/0501272}. The resulting variation of the mixing angles
with $\epsilon$ is shown in Fig.~\ref{antusch-plot}. The broadening of the
allowed regions of the atmospheric and reactor mixing angles is caused by the
phase dependence of $\theta_{23}$ and $\theta_{13}$,
cf. Eqs.~(\ref{t23},\ref{t13}). In the case of the solar mixing angle, our
analytic formula, Eq.~\eqref{t12}, does not show any phase dependence to
second order in $\epsilon$, so the significantly smaller broadening of the
allowed region of $\theta_{12}$ is an effect of third order in $\epsilon$. 

We can express our results in terms of the deviation parameters $s$, $a$
and $r$ as defined in~\cite{King:2007pr},  
\be
 \sin \theta_{12} = \frac{1}{\sqrt{3}}(1+s)\ , \qquad
\sin \theta_{23}  =  \frac{1}{\sqrt{2}}(1+a)\  , \qquad 
 \sin \theta_{13} =   \frac{r}{\sqrt{2}}\ .
\label{rsa-2}
\ee
Comparing these definitions with our second order expressions in
Eqs.~(\ref{t12}-\ref{al3}) we find
\be
s=-\mbox{$\frac{\epsilon^2}{6}$} \ , \quad ~
a=\mbox{$\frac{\epsilon^2}{3}$} \, k \cos(\alpha_{23}+\delta)  \ ,
\quad ~ 
r=\epsilon + \mbox{$\frac{\epsilon^2}{3}$} \,k
\cos(\alpha_{23}+2\delta) \ ,
\ee
where we have replaced the original phase parameters by the two physical phases
$\delta$ and $\alpha_{23}$, keeping only terms up to second order in
$\epsilon$. These relations can be combined to yield the second order sum
rules
\be
s=-\mbox{$\frac{r^2}{6}$} \ , \qquad 
a=\mbox{$\frac{r^2}{3}$} \, k \cos(\alpha_{23}+\delta) 
= \mbox{$\frac{r}{3}$} \, \mbox{$\frac{m_2}{m_3}$} \, \cos(\alpha_{23}+\delta)  \ .
\ee
Using these sum rules and the second order expansion of the PMNS matrix given
in~\cite{King:2007pr}, we can rewrite the mixing matrix of PCSD in terms of
the reactor deviation parameter~$r$ and the parameter 
$\kappa = k \cos(\alpha_{23}+\delta)$,

\be
U_{\mathrm{PMNS}} =
\begin{pmatrix}
  \sqrt{\frac{2}{3}} \left(1-\frac{r^2}{6} \right) 
& \frac{1}{\sqrt{3}} \left(1-\frac{5r^2}{12}  \right) 
& \frac{r}{\sqrt{2}} e^{-i \delta }  \\
 -\frac{1}{\sqrt{6}} \left(1 +r e^{i \delta } - \frac{1+2\kappa}{6} r^2  \right) 
& \phantom{-}\frac{1}{\sqrt{3}} \left(1-\frac{r}{2} e^{i \delta } + \frac{1-4 \kappa}{12}  r^2 \right) 
& \frac{1}{\sqrt{2}} \left(1 - \frac{3-4\kappa}{12} r^2   \right)\\
\phantom{-} \frac{1}{\sqrt{6}} \left(1- r  e^{i \delta } - \frac{1-2\kappa}{6}  r^2  \right) 
& -\frac{1}{\sqrt{3}} \left(1 + \frac{r}{2} e^{i \delta } +\frac{1+4\kappa}{12}  r^2  \right) 
& \frac{1}{\sqrt{2}}\left(  1 -\frac{3+4\kappa}{12}  r^2   \right) 
\end{pmatrix} P \, .
\label{PMNS-PCSD}
\ee
The comparison with the second order expansion of the TBR mixing matrix in
Eq.~\eqref{TBR}, where $s=a=0$,
\bea
U_{\mathrm{TBR}} =
\left( \begin{array}{ccc}
\sqrt{\frac{2}{3}} \left(1-\frac{r2}{4}\right) & \frac{1}{\sqrt{3}}
\left(1-\frac{r2}{4}\right)& \frac{r}{\sqrt{2}}  e^{-i\delta} \\
-\frac{1}{\sqrt{6}}(1+ r e^{i\delta})
& \phantom{-}\frac{1}{\sqrt{3}}(1- \frac{r}{{2}}e^{i\delta})
& \frac{1}{\sqrt{2}} \left(1-\frac{r2}{4}\right)\\
\phantom{-}\frac{1}{\sqrt{6}}(1- r e^{i\delta})
& -\frac{1}{\sqrt{3}}(1+ \frac{r}{{2}} e^{i\delta})
& \frac{1}{\sqrt{2}}\left(1-\frac{r2}{4}\right)
\end{array}
\right)P\ .
\label{TBR-second}
\eea
illustrates how accurately TBR mixing is achieved in models of PCSD.

It is also worth noting that the columns of the Dirac mass matrix 
in Eq.\eqref{mR} are not quite proportional to the columns of the 
PMNS matrix in Eq.\eqref{PMNS-PCSD}, so that form dominance \cite{arXiv:0903.0125}
is violated at order $O(r)$. This implies that leptogenesis is non-zero \cite{arXiv:1004.3756},
even in the absence of renormalisation group corrections \cite{arXiv:1110.5676}.




\section{\label{sec-concl}Conclusions}
\cleqn

Recent results from T2K, MINOS and Double CHOOZ all indicate a sizeable reactor angle $\theta_{13}$
which would rule out {\em conventional} tri-bimaximal lepton mixing.
However, it is possible to maintain the tri-bimaximal solar and atmospheric mixing angle predictions, 
 $\theta_{12}\approx 35^\circ$, $\theta_{23}\approx 45^\circ$ even for a quite sizeable 
reactor angle such as $\theta_{13}\approx ~8^\circ$, using an ansatz
called tri-bimaximal-reactor (TBR) mixing in Eq.~\eqref{TBR} proposed by one of us some time ago,
along with the neutrino mass matrix in Eq.~\eqref{quadratic} arising from PCSD \cite{King:2009qt}.

In this paper we have proposed the first explicit $A_4$ model of leptons 
based on the type~I seesaw mechanism at both the effective and the 
renormalisable level which, together with vacuum alignment, 
leads to the desired form of neutrino mass matrix. 
After performing an analytic diagonalisation of the neutrino mass matrix 
to second order, we find that the coefficients of the second order terms are suppressed, 
making TBR mixing surprisingly accurate for the renormalisable $A_4$ model. The analytic results are confirmed
by the numerical results in Fig.~\ref{antusch-plot} which illustrates the stability of the
atmospheric and solar angles as the reactor angle is switched on.

It is worth emphasising that the $A_4$ models are {\em indirect} models involving
the quadratic appearance of misaligned flavons as in Eq.~\eqref{quadratic}. 
Moreover there is no simple Klein symmetry that is respected by this mass
matrix; in fact, formally, the generators of the Klein symmetry
involve the real order one parameter $k = \frac{|m^0_2|}{\epsilon |m^0_3|}$ of
Eq.~\eqref{k-param}, thus proving the absence of a {\em direct} link between the
$A_4$ family symmetry and the Klein symmetry of the neutrino mass matrix. 
Nevertheless, the alignments of the flavons can readily originate from a simple discrete
symmetry such as $A_4$, which is broken in a rather complicated way in the
neutrino sector, although the charged lepton sector is diagonal in the model
of leptons considered here.  

In a more complete $A_4$ family unified model, for example
based on $SU(5)$, one would expect the charged lepton sector to be related to the off-diagonal quark mass matrices, resulting in additional charged lepton corrections to the lepton mixing angles, together with renormalisation group (RG) corrections, as discussed in \cite{arXiv:0810.3863}. However it is worth emphasising that, typically, such corrections to mixing angles are not expected to exceed about 3$^\circ$ from charged lepton corrections
and about 1$^\circ$ from RG corrections \cite{arXiv:0810.3863}.
Such corrections are not sufficient to account for the observed reactor angle, although they may affect the predictions of TBR mixing discussed here, which strictly apply to the unrenormalised neutrino mixing angles only.




\section*{Acknowledgements}

We are very grateful to Stefan Antusch for generously creating and donating
the Figure, as well as for helpful correspondence. The authors acknowledge support
from the STFC Rolling Grant No. ST/G000557/1 and partial support by the EU ITN
grant UNILHC 237920  (Unification in the LHC era). 




\section*{Appendix}

\begin{appendix}

\section{\label{sec-withM1}Analytic diagonalisation of the PCSD neutrino mass matrix with $\bs{m_1^0 \neq  0}$ to second order}
\cleqn
In this appendix we present the results of the mixing angles and phases for
the more general case with {\it three} non-zero mass parameters $m_i^0$. 
The starting point is the PCSD 
form of neutrino mass matrix with non-vanishing $m_1^0$, namely,
\be
m_\nu ~=~ 
\frac{m^0_1}{6} \begin{pmatrix}
4&-2&2 \\
-2 &1&-1\\
2 &-1&1
\end{pmatrix}
+
\frac{m^0_2}{3} \begin{pmatrix}
1&1&-1 \\
1&1&-1\\
-1 &-1&1
\end{pmatrix}
+
\frac{m^0_3}{2} \begin{pmatrix}
\varepsilon^2&\varepsilon&\varepsilon \\
\varepsilon&1&1\\
\varepsilon &1&1
\end{pmatrix}.
\label{m-analytic-m1}
\ee
Clearly, in the limit $m_1^0=0$, this reduces to the PCSD form of neutrino mass matrix considered 
in Eq.~\eqref{quadratic} and Section \ref{sec-tbr}. 
Adopting the notation of Section~\ref{sec-tbr} and assuming the hierarchy
\be
|m^0_1|~=~k_1 \,\epsilon \,|m^0_3| \ , \qquad 
|m^0_2|~=~k_2 \,\epsilon \,|m^0_3| \ ,
\ee
with $k_i \in \mathbb R$ being coefficients of order one or smaller, we obtain
to second order in $\epsilon$,
\bea
\theta_{12}&=& \arcsin \mbox{$\frac{1}{\sqrt{3}}$} -  \mbox{$\frac{k_2^2
    +k_1^2+2  k_2k_1 \cos (\alpha^0_2-\alpha^0_1)}{6 \sqrt{2}(k_2^2-k_1^2)}$}\, 
\epsilon^2 \ ,  \\
\theta_{23}&=& \mbox{$\frac{\pi}{4}$}+ \mbox{$\frac{1}{3}$} 
\big[ k_2 \cos(\alpha^0_2-\alpha^0_3+\delta^0) -k_1  \cos(\alpha^0_1-\alpha^0_3+\delta^0)\big] \,
\epsilon^2 \ , \\[1.5mm]
\theta_{13}&=& \mbox{$\frac{\epsilon}{\sqrt{2}}$} +
\mbox{$\frac{1}{3\sqrt{2}}$} 
\big[ k_2 \cos(\alpha^0_2-\alpha^0_3+2\delta^0) +2k_1 \cos(\alpha^0_1-\alpha^0_3+2\delta^0)\big] \,
\epsilon^2 \ , \\[1.5mm]
\delta &=& \delta^0 - \mbox{$\frac{1}{3}$}
\big[ k_2 \sin(\alpha^0_2-\alpha^0_3+2\delta^0) +2 \,k_1 \sin(\alpha^0_1-\alpha^0_3+2 \delta^0)\big] \,
\epsilon \ , \\[1mm]
\alpha_{21}&=&   \alpha^0_2-\alpha^0_1
 +  \mbox{$\frac{k_2 k_1 \sin(\alpha^0_2-\alpha^0_1)}{3(k_2^2-k_1^2)}$}\, 
\epsilon^2  \ ,\\[1mm]
\alpha_{31}&=& 
\alpha^0_3-\alpha^0_1
+  \mbox{$\frac{2 k_2 k_1 \sin(\alpha^0_2-\alpha^0_1)}{3(k_2^2-k_1^2)}$}\, 
\epsilon^2  \ , \\[1mm]
\delta_e &=&  \mbox{$\frac{\alpha^0_1}{2}$} 
+  \mbox{$\frac{k_2k_1 \sin(\alpha^0_2-\alpha^0_1)}{6(k_2^2-k_1^2)}$} \, \epsilon^2 \ ,\\[1mm]
\delta_\mu &=&  \mbox{$\frac{\alpha^0_1}{2}$} +  \mbox{$\frac{1}{3}$} 
\big[ k_2 \sin(\alpha^0_2-\alpha^0_3+\delta^0)-
\mbox{$\frac{k_2k_1 \sin(\alpha^0_2-\alpha^0_1)}{k_2^2-k_1^2}$}
-k_1 \sin(\alpha^0_1-\alpha^0_3+\delta^0)
\big] \, \epsilon^2  ,\\[1mm]
\delta_\tau &=&  \mbox{$\frac{\alpha^0_1}{2}$} -  \mbox{$\frac{1}{3}$} 
\big[ k_2 \sin(\alpha^0_2-\alpha^0_3+\delta^0)+
 \mbox{$\frac{k_2k_1 \sin(\alpha^0_2-\alpha^0_1)}{k_2^2-k_1^2}$}
-k_1 \sin(\alpha^0_1-\alpha^0_3+\delta^0)
\big] \, \epsilon^2 .~~~~~~~
\eea
The resulting neutrino masses read
\be
m_\nu^{\mathrm{diag}} 
~=~ 
\text{diag} \left(m_1\,,\,m_2\,,\,m_3\right) 
~=~ 
\text{diag} \left(k_1 \,\epsilon \,,\, k_2 \,\epsilon \,,\, 1+
\mbox{$\frac{\epsilon^2}{2}$}\right)  \, |m^0_3|\ .
\ee
Note that $m_1$ can be of same order in $\epsilon$ as $m_2$ without changing
the mixing matrix to first order in $\epsilon$. That is even with $k_1$ being
of order one, we still find TBR mixing. 
Considering, however, the case where $|m^0_1|$ is additionally suppressed by
one or more powers of $\epsilon$, i.e. $k_1 \sim \mathcal O(\leq \epsilon)$,
the above expressions simplify considerably. Then, in fact, the $k_1$
dependence drops out completely to second order in $\epsilon$, and we recover the
expressions of Section~\ref{sec-tbr}, with $k_2$ corresponding to $k$ and
$\alpha^0_1$ set to zero. 

\end{appendix}




\end{document}